\documentclass[final,5p,times,twocolumn]{elsarticle}

\usepackage{color,amssymb}

\usepackage{hyperref}

\journal{Physics Letters B}

\newcommand{\black}{\color{black}}
\newcommand{\red}{\black}

\begin{document}

\begin{frontmatter}


\title{
Energy-dependent flavor ratios, cascade/track spectrum tension and high-energy neutrinos from magnetospheres of supermassive black holes
}


\author[MSU]{Kirill Riabtsev}
\ead{riabtcev.ki19@physics.msu.ru}
\affiliation[MSU]{organization={Physics Department, M.V.~Lomonosov Moscow State University},
            addressline={1-2 Leninskie Gory}, 
            postcode={119991}, 
            city={Moscow},
            country={Russia}}

\author[INR]{Sergey Troitsky}

\affiliation[INR]{organization={Institute for Nuclear Research of the Russian Academy of Sciences},
            addressline={60th October Anniversary Prospect 7A}, 
            postcode={117312}, 
            state={Moscow},
            country={Russia}}
\date{\rm INR--TH-2022-009}
\begin{abstract}
The IceCube neutrino observatory measures the diffuse flux of high-energy astrophysical neutrinos by means of various techniques, and there exists a mild tension between spectra obtained in different analyses. The spectrum derived from reconstruction of muon tracks is harder than that from cascades, dominated by electron and tau neutrinos. If confirmed, this tension may provide a clue to the origin of these neutrinos, which remains uncertain. Here we investigate the possibility that this tension may be caused by the change of the flavor content of astrophysical neutrinos with energy. We assume that at higher energies, the flux contains more muon neutrinos than expected in the usually assumed flavor equipartition. This may happen if the neutrinos are produced in regions of the magnetic field so strong that muons, born in pi-meson decays, cool by synchrotron radiation faster than decay. The magnetic field of $\sim 10^4$~G is required for this mechanism to be relevant for the IceCube results. We note that these field values are reachable in the immediate vicinity of supermassive black holes in active galactic nuclei and present a working toy model of the population of these potential neutrino sources. While this model predicts the required flavor ratios and describes the high-energy spectrum, it needs an additional component to explain the observed neutrino flux at lower energies.
\end{abstract}




\end{frontmatter}


\section{Introduction}
\label{sec:sample1}
The origin of high-energy astrophysical neutrinos is one of the most intriguing questions of modern astroparticle physics (see e.g.\ Ref.~\cite{ST-UFN} for a recent review). The existence of these energetic (TeV to PeV) particles born outside the terrestrial atmosphere was established by the IceCube experiment \cite{IceCubeFirstPeV,IceCubeFirst26} and starts to be confirmed by ANTARES \cite{ANTARES} and Baikal-GVD \cite{BaikalSymmetry}, but astrophysical sources and production mechanisms of the neutrinos are uncertain.  

Present experimental data on high-energy neutrinos are dominated by observations with IceCube, a neutrino telescope  at the South Pole in Antarctica. Two principal groups of detected events are cascades and tracks, and they give complementary information about the astrophysical neutrino flux. Most of cascade events are caused by electron, $\nu_e$, or tau, $\nu_\tau$, neutrinos, while most of the track events are caused by muon neutrinos, $\nu_\mu$. They also have different distributions in energies and arrival directions, as well as in the accuracy of measurement of these observables. 

Neutrinos of these energies can be produced in decays of secondary mesons born in hadronic interactions. Assuming decays of $\pi$ mesons in the source and taking into account neutrino oscillations on their way from the source to the observer, one expects roughly equal flavor ratio of fluxes at the detector, $\nu_e:\nu_\mu:\nu_\tau=1:1:1$, see e.g.\ Ref.~\cite{VissaniUniverse} for an updated calculation. Since this scenario is the most general, the flavor equality is assumed in the reconstruction of the observed neutrino spectra. 

There exists a mild tension between parameters of the spectra reconstructed from IceCube cascade and track observations, discussed e.g.\ in Refs.~\cite{2compViss2016anomaly,IceCube:HESE2020,ST-UFN}. Deviations from the assumed power-law shape, caused for instance by the presence of two distinct populations of sources contributing to the total observed neutrino flux, may explain this discrepancy \cite{Kohta699,2compChen2014,2compViss2016,2compNerSem2016}. 

In the present Letter, we discuss the possibility that the tension appears due to a deviation from the assumed flavor content. In Sec.~\ref{sec:spectra}, we use the most recent IceCube results on the spectra of $\nu_\mu$ and $\nu_e+\nu_\tau$ from independent analyses to calculate the observed flavor ratio as a function of energy and to see that a change in the flavor content at the energy $\sim 100$~TeV is plausible. Then, in Sec.~\ref{sec:mu-damp}, we consider a possible mechanism predicting the deviation from flavor equipartition, known as the muon damping, and estimate the magnetic field in the source required for the transition to this regime around $\sim 100$~TeV. The required values of $\sim 10^4$~G are not very common in astrophysical sources, but may be present in the immediate vicinity of supermassive black holes in active galactic nuclei (AGN), and we construct and study quantitatively a toy model of a population of these sources in Sec.~\ref{sec:BH}. Section~\ref{sec:concl} presents our brief conclusions. 


\section{Comparison of IceCube $\nu_\mu$ and $\nu_e+\nu_\tau$ fluxes}
\label{sec:spectra}
More than 90\% of neutrino events in the track sample are caused by $\nu_\mu$ \cite{IceCube:90percentMu}, so we use the flux reconstructed from track events as a proxy for the $\nu_\mu$ flux. Note that the energy of an individual neutrino producing the track is determined with very low precision, so the spectrum is reconstructed only under assumptions of a particular model. The best-fit reconstructed spectrum is provided by a power law, and we use its parameters from the most recent published IceCube track analysis \cite{IceCube:mu2021}.

Energies of neutrinos causing cascade events are determined much more precisely, thus allowing one to reconstruct a binned differential spectrum. Moreover, particular techniques have been applied by IceCube to separate a clean subsample of cascades dominated by $\nu_e$ and $\nu_\tau$ \cite{IceCube:e-tau-flux,Halzen:2019lxc}. We use these data to determine the ratio of the mean flux of $\nu_e$ and $\nu_\tau$ to that of $\nu_\mu$ in the energy bins for which both analyses are valid, that is between $\sim 15$~TeV and $\sim 5$~PeV.

\begin{figure}
\centering
\includegraphics[width=\linewidth]{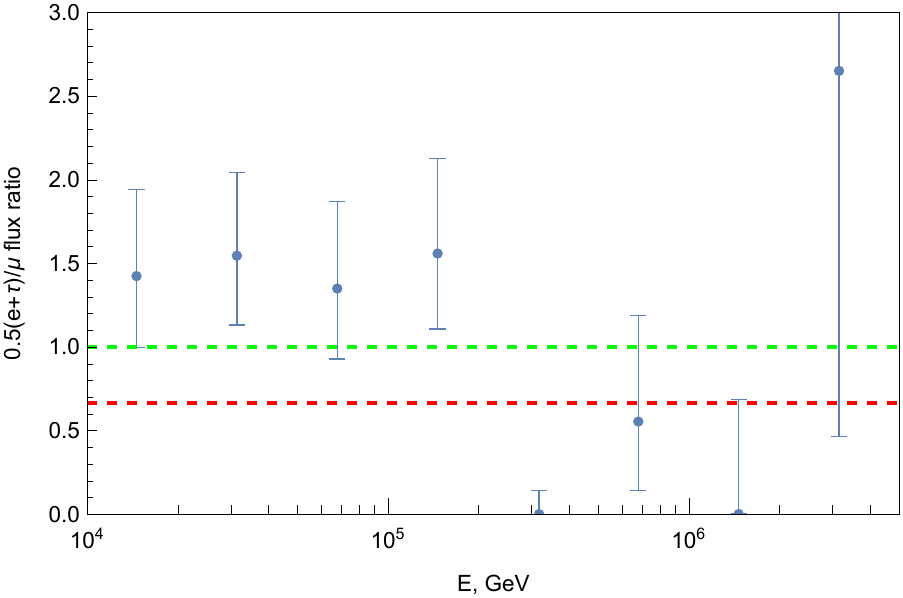}
\caption{\label{fig:ratio}
The ratio of observed fluxes of neutrinos of different flavours as a function of energy. The dashed green line represents flavor equipartition, while the dashed red line corresponds to 2/3, predicted in the muon damping case. See the text for details on the data points.
}
\end{figure}

Figure~\ref{fig:ratio} 
presents the energy-dependent estimates of the flavour ratio obtained in this way. For comparison, lines of constant ratios, 1 predicted in the standard mechanism with $\pi$-meson decays and $2/3$ predicted in its variation with ``damped muons'', discussed bellow in Sec.~\ref{sec:mu-damp}, are also shown. 
One can see that the change of the flavor composition at the neutrino energies $\sim 200$~TeV describes the data better than a constant flavor ratio.
Below we present more detailed calculations in terms of separate $\nu_\mu$ and $\nu_e+\nu_\tau$ spectra in the frameworks of a particular quantitative model.


\section{Transition to the muon-damped regime}
\label{sec:mu-damp}
Assume that the standard process of hadronic interactions and subsequent decay of $\pi$ mesons, giving rise to high-energy neutrinos in the most common scenario, takes place in a region of strong magnetic field. If the synchrotron radiation loss time scale of the muons is shorter than the decay time scale, the muons lose much of their energy before the decay. Hence, only primary neutrinos from pion decays contribute to the high-energy flux, while neutrinos from decays of these secondary muons show up at considerably lower energies \cite{Rachen-muoncooling,Beacom-mudamp,Waxman-mudamp}. This regime of high-energy neutrino production has been called ``muon damp'' and was recognized in a number of works discussing the flavor content of IceCube neutrinos, see e.g.\ Refs.~\cite{Winter-Hillas2010,Tamborra-magnetometers}. The flavor content of high-energy neutrinos is different in this case from the one normally assumed: with the account of neutrino oscillations on their way to the observer it becomes $\nu_e:\nu_\mu:\nu_\tau\approx 1:2:2$, see e.g.\ Refs.~\cite{Beacom-mudamp,Weiler-ratios}. 

\subsection{Required magnetic fields}
Though detailed calculations of the spectra of neutrinos of various flavors require modelling of actual conditions in the source, it is easy to obtain an estimate of the magnetic-field strength at which the muon damp starts to dominate: it is given by the condition $t_{\rm sync}<t_{\rm decay}$, where $t_{\rm sync}=9m_\mu^4/(4e^4 B^2 \mathcal{E}_\mu)$ is the muon synchrotron cooling time, $t_{\rm decay}= \mathcal{E}_\mu \tau_\mu/m_\mu$ is the muon lifetime; $m_\mu$, $\mathcal{E}_\mu$, $\tau_\mu$ and $e$ are the muon mass, energy, lifetime at rest and charge, respectively, and $B$ is the external magnetic field. We obtain the muon-damp condition
\begin{equation}
\label{Eq:BfromEmuon-damp}
    B\gtrsim \frac{3}{2 e^2} m_\mu^{5/2} \mathcal{E}_\mu^{-1} \tau_\mu^{-1/2} \approx 
    2\times 10^4~\mbox{G}\,\left(\frac{\mathcal{E}_\nu}{\mbox{100~TeV}}\right)^{-1},
\end{equation}
where we used the pion-decay kinematics to approximately relate the muon and neutrino energies as $\mathcal{E}_\mu\approx 3\mathcal{E}_\nu$, see e.g.\ Ref.~\cite{Lipari:2007su}.

\subsection{Candidate sources}
The magnetic fields of order of tens kiloGauss, implied by Eq.~(\ref{Eq:BfromEmuon-damp}), are not common in potential astrophysical neutrino sources, see e.g.\ Ref.~\cite{PT-Hillas} for a review of physical conditions in various astrophysical accelerators. Among extragalactic sources, only gamma-ray bursts (GRBs) and magnetospheres of supermassive black holes (SMBH) are believed to host these high fields.

Being in principle plausible candidates for particle acceleration, GRBs as high-energy neutrino sources are disfavoured because of the lack of directional and temporal correlations between the neutrino arrivals and known GRBs \cite{NoGRB}. A possible way to overcome this constraint is to invoke the population of hidden GRBs, e.g.\ those with so-called choked jets \cite{Meszaros:chocked}. However, the magnetic fields in the neutrino production regions are poorly known for this scenario, and more complicated mechanisms affecting the flavor composition of produced neutrinos may be at work there, see e.g.\ Refs.~\cite{Meszaros:chocked,Kohta891}. 

The magnetic fields near SMBHs are also poorly constrained observationally, so we use the benchmark Shakura-Sunyaev \cite{ShakuraSunyaev} estimate of
\begin{equation}
    B \sim 10^4~{\rm G} \left( \frac{M}{10^8 M_\odot}\right)^{-1/2},
    \label{eq:Shakura-Syunyaev}
\end{equation}
where $M_\odot$ denotes the Solar mass and $M$ os the SMBH mass. The immediate vicinity of a supermassive black hole with mass $M\sim 10^8 M_\odot$, typical for AGN, becomes a plausible place of the neutrino production. \red Numerous models of neutrino production in AGN have been proposed, see e.g.\ reviews \cite{ST-UFN,Boettcher-rev,Cerruti-rev,MuraseSteckerRev} and many references therein, including but not limited to \cite{BerezinskyNeutrino77,Eichler1979,BerezGinzb,Sikora1990,Stecker:1991vm,AtoyanDermer,Dermer:AGN,NeronovWhich,Stecker:AGN,Kalashev:AGN,MuraseCorona1,MuraseCorona2,InoueCorona1,InoueCorona2}, but only some of them put the neutrino production region very close to SMBH. \black  One possibility is that protons may be accelerated in the vacuum gap of the black-hole magnetosphere up to the required energies \cite{gap1,PtitsynaNeronov-gap}, and then interact with ambient photons to produce $\pi$ mesons, cf.\ e.g.\ Refs.~\cite{Stecker:AGN,Dermer:AGN,Kalashev:AGN}. The neutrino flux enhancement due to relativistic jets pointing at the observer may then explain the observed correlations of detected neutrinos with radio blazars, putting the neutrino production region to the very central parts of AGN \cite{Plavin1,Plavin2,Hovatta:2020lor,Plavin-ICRC}, see also Refs.~\cite{Resconi,vanVelzen:2021zsm,Kun:2022khf}. In what follows, we discuss the neutrino production in the SMBH magnetospheres in more detail.

\section{Black-hole magnetospheres in AGN as neutrino sources: a toy model}
\label{sec:BH}
We now present a toy model demonstrating that the IceCube neutrino spectra, both for $\nu_\mu+\nu_\tau$ and for $\nu_e$, can indeed be explained with the transition to the muon-damp regime if the neutrinos originate from immediate vicinity of SMBHs. \red This toy model serves only to demonstrate that there exist astrophysical sources where reasonable neutrino fluxes may be generated in the muon-damped regime at the energies relevant for IceCube observations. Note that previous studies, e.g.\ Ref.~\cite{Tamborra-magnetometers}, considered energies beyond the present reach of IceCube. \black

We assume that relativistic protons are accelerated in the vacuum gap of the SMBH magnetosphere \cite{gap1,PtitsynaNeronov-gap}. This mechanism produces, to the first approximation, a monochromatic proton spectrum, with the energy limited either by the gap size or by losses. This energy is however different for different sources and, since the number of the sources is large, together they produce a particular continuous spectrum of protons, see Ref.~\cite{population}. We use updated estimates of the achievable proton energies from Ref.~\cite{PtitsynaNeronov-gap} and express the proton energy in the source frame, $\mathcal{E}'_p$, through the SMBH mass $M$ and the Eddington ratio $\lambda$, which determine the magnetic field, by Eq.~(\ref{eq:Shakura-Syunyaev}), and the luminosity. We further use the benchmark relation $\mathcal{E}'_\nu\approx\mathcal{E}'_p/20$, see e.g.\ Ref.~\cite{VissaniUniverse}, relevant both for $pp$ processes and for $p\gamma$ interactions in the $\Delta$ resonance approximation. This determines the contribution of a particular SMBH. \red Note that for the monochromatic proton spectrum, the neutrino spectrum is also essentially monochromatic, both for the $pp$ and $p\gamma$ cases. \black Next, we sum these \red individual SMBH \black contributions to obtain the neutrino spectrum.

\begin{figure}
\centering
\includegraphics[width=\linewidth]{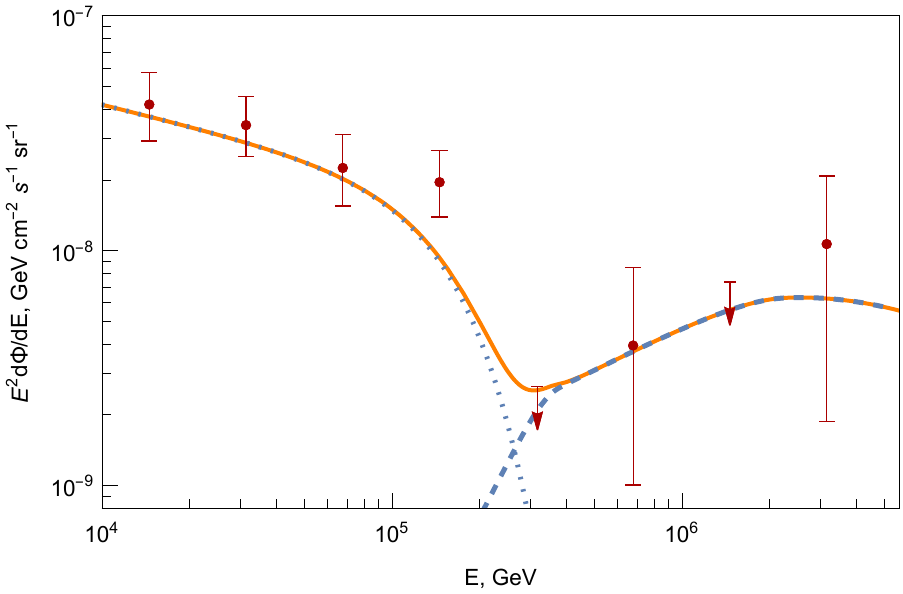}
\caption{\label{fig:model-etau}
Comparison of the $\nu_e+\nu_\tau$ spectrum determined by IceCube \cite{Halzen:2019lxc} and predictions of the toy model discussed in the text.
}
\end{figure}

\begin{figure}
\centering
\includegraphics[width=\linewidth]{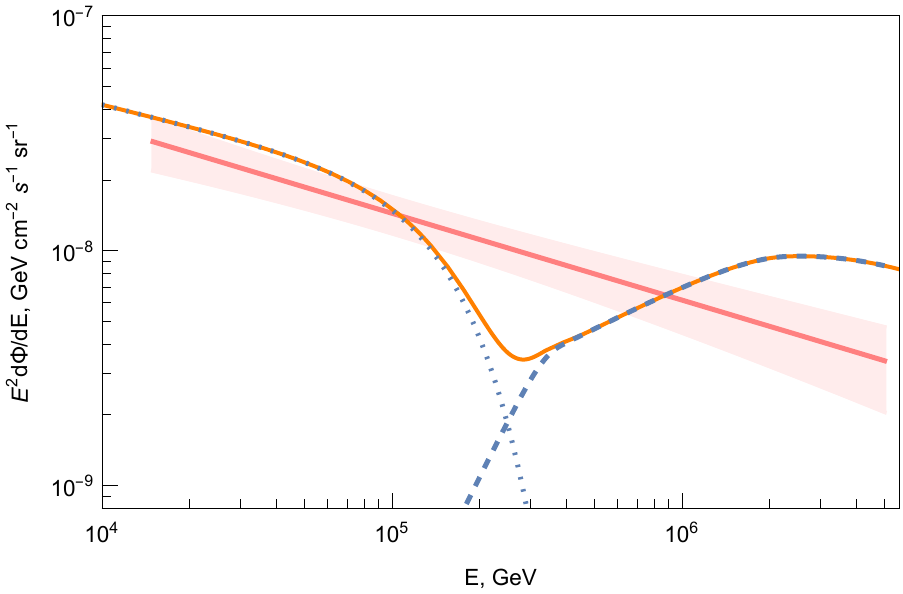}
\caption{\label{fig:model-mu}
Comparison of the $\nu_\mu$ spectrum determined by IceCube \cite{IceCube:mu2021} and predictions of the toy model discussed in the text.
}
\end{figure}

To this end, we integrate over the distributions of SMBHs in $M$,  $\lambda$ and over redshift $z$ and obtain the neutrino spectrum in terms of the observed neutrino energies $\mathcal{E}_\nu$. The normalization of the spectrum is kept arbitrary at this step. We benefit from a recent study \cite{shadows}, where a working up-to-date model of both the redshift-dependent SMBH mass function and the redshift-dependent Eddington ratio distribution is compiled, based on Refs.~\cite{BHMF} and \cite{ERF}, respectively. We take into account the contribution of active galaxies only, which we define as those with $\lambda>0.01$, as it is customary in the SMBH population studies. Whether the contribution of a particular SMBH is in the muon-damp regime depends on the fulfillment of the condition (\ref{Eq:BfromEmuon-damp}), which is determined in turn by its $M$ and $\lambda$. We calculate the resulting observed flux of neutrinos of various flavors and normalize the all-flavor sum to match the IceCube observations, see dashed lines in Figs.~\ref{fig:model-etau} and \ref{fig:model-mu}, respectively.
\begin{figure}
\centering
\includegraphics[width=\linewidth]{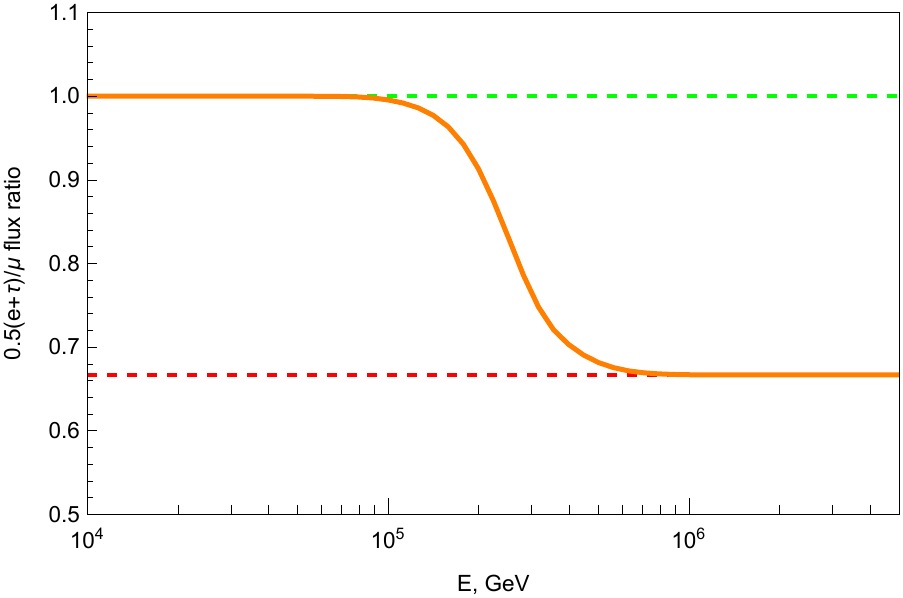}
\caption{\label{fig:model-ratio}
Energy-dependent neutrino flavor ratio predicted by the toy model discussed in the text (solid curve). The dashed green line represents flavor equipartition, while the dashed red line corresponds to 2/3, predicted in the muon damping case.
}
\end{figure}

As it is common in models of neutrino production close to SMBHs, the contribution from this mechanism explains the observed flux only at energies $\mathcal{E}_\nu \gtrsim 100$~TeV. To obtain the entire observed astrophysical spectrum, we add a second component at lower energies, e.g.\ \cite{Kohta491,Kohta740}, which we parametrize as a power law with double exponential cutoff adjusted to fit, together with the SMBH contribution, the observed spectrum of $\nu_e+\nu_\tau$. This component is shown by dotted lines in Figs.\ \ref{fig:model-etau} and \ref{fig:model-mu}. Whatever particular astrophysical sources contribute to this component, they are not in the muon-damp regime because they do not have sufficiently strong magnetic fields. \red While this additional component may be related to various objects, we note that most of the proposed extragalactic high-energy neutrino sources, including blazars, tidal disruption events, Seyfert galaxies etc., all host SMBHs and therefore can contribute to the muon-damped regime described by our toy model. \black The resulting energy-dependent flavor ratio is shown in Fig.~\ref{fig:model-ratio}.

This function, derived in our toy model, improves the fit of the data in Fig.~\ref{fig:ratio} by $\Delta(\chi^2/\mbox{d.o.f.})\approx 3.0$ compared to the flavor equipartition. Accumulation of the data from IceCube and complementary new experiments, Baikal-GVD \cite{Baikal-GVD:2021zsq}, KM3NeT \cite{KM3NeT}, IceCubeGEN2 \cite{IceCube-Gen2:2020qha}, P-ONE \cite{P-ONE} and TAMBO \cite{TAMBO}, would improve the discrimination: following Ref.~\cite{flavor:future} and allowing in addition for data taking in incomplete configurations of future detectors, we find that $\Delta(\chi^2/\mbox{d.o.f.})\sim 5$ might be reached in six years. 

\section{Conclusions}
\label{sec:concl}
It is presently unclear, whether the long-standing tension between neutrino energy spectra reconstructed by IceCube from cascades and tracks is caused by systematic uncertainties or by physical reasons. In the latter case, it may serve as a clue to determination of mechanisms and sources responsible for production of high-energy astrophysical neutrinos. 

We assumed that the reason for the discrepancy is related to the change of flavor composition of high-energy neutrinos with energy and demonstrated that this agrees with observational data, if the standard flavor equipartition switches to the flavor content predicted in the muon-damp regime at energies above $\sim 100$~TeV. Then we estimated the magnetic field at sources required for this switch and found that it is of order $\sim 10^4$~G. We noted that fields of this scale may exist in immediate environments of supermassive black holes and constructed a quantitative toy model of the population of these high-energy neutrino sources. We found that this model describes well the spectra of both $\nu_\mu$ and $\nu_e+\nu_\tau$ provided an additional component with standard flavor content is added at low energies.

The idea to infer the magnetic field, and hence to constrain the sources of high-energy neutrinos, from observation of the change in the flavor composition, was discussed in Ref.~\cite{Tamborra-magnetometers}. However, that work did not make use of IceCube track data and therefore did not address the cascade/track tension as a potential signal of the change of the flavor content with energy. Our estimates of the magnetic field in the neutrino sources agree with upper limits obtained in Ref.~\cite{Tamborra-magnetometers}.

Measurements of the flavor content of astrophysical neutrinos serve as a useful tool to constrain the neutrino origin \cite{flavor:tool}. While precise measurement of the flavor content at the source would require much better statistics than currently available \cite{flavor:future}, the scenario we discuss here might be tested with future data in less than ten years. 

We are indebted to T.~Dzhatdoev and K.~Murase for useful comments. This work is supported by the RF Ministry of science and higher education under the contract 075-15-2020-778. 


 \bibliographystyle{elsarticle-num} 
 \bibliography{flav}

\end{document}